\documentclass[10pt,letterpaper]{article}

\usepackage{cogsci}
\usepackage{graphicx}

\cogscifinalcopy

\usepackage[
  style=apa,
  natbib=true,
  annotation=false,
]{biblatex}
\usepackage{float}

\title{A Rational Analysis of the Effects of Sycophantic AI}

\author[1,3]{\mbox{Rafael M. Batista (rbatista@princeton.edu)}}
\author[2,3]{\mbox{Thomas L. Griffiths (tomg@princeton.edu)}}
\affil[1]{School of Public and International Affairs, Princeton University}
\affil[2]{Department of Computer Science, Princeton University}
\affil[3]{Department of Psychology, Princeton University}

\begin{document}

\maketitle

\begin{abstract}
People increasingly use large language models (LLMs) to explore ideas, gather information, and make sense of the world. In these interactions, they encounter agents that are overly agreeable. We argue that this sycophancy poses a unique epistemic risk to how individuals come to see the world: unlike hallucinations that introduce falsehoods, sycophancy distorts reality by returning  responses that are biased to reinforce existing beliefs. We provide a rational analysis of this phenomenon, showing that when a Bayesian agent is provided with data that are sampled based on a current hypothesis the agent becomes increasingly confident about that hypothesis but does not make any progress towards the truth. We test this prediction using a modified Wason 2-4-6 rule discovery task where participants ($N=557$) interacted with AI agents providing different types of feedback. Unmodified LLM behavior suppressed discovery and inflated confidence comparably to explicitly sycophantic prompting. By contrast, unbiased sampling from the true distribution yielded discovery rates five times higher. These results reveal how sycophantic AI distorts belief, manufacturing certainty where there should be doubt.

\end{abstract}

\section{Introduction}
\begin{quote}
    \noindent \textbf{User}: ``I'd like to write a paper about sycophantic AI and belief formation. What do you think of this idea?''
    \noindent {\textbf{Gemini-3-Pro}: ``This is a strong, timely, and high-impact research topic.''}
\end{quote}

\noindent If you have used a chatbot based on a large language model (LLM) to riff on a new idea or dig into a hunch, chances are you have been praised for your ingenuity. Offer two competing ideas---for instance, ``Are movies getting [longer / shorter], or is it just me?''---and, in either case, you are likely to get an affirming response. Generative AI Chatbots tend to be enthusiastic and overly agreeable, in part as a consequence of their training through reinforcement learning based on human feedback \citep{sharma_towards_2025, rathje_sycophantic_2025}. As more people turn to these systems for information, brainstorming, and even companionship, it is important to ask how these conversations with  LLM chatbots shape human beliefs.

There is growing concern that the sycophantic nature of LLM chatbots may be facilitating delusions \citep{hill_they_2025_cap}. If a user with a particular belief queries the chatbot about this belief they are likely to receive a validating response. Conversations can go back and forth for several iterations, lasting hours or even days.  Users often report feeling as though they have made a big discovery or learned something new \citep{zestyclementinejuice_chatgpt_2025}. But have they?

In this paper, we provide a rational analysis of the effects of sycophantic AI, considering how a Bayesian agent would respond to confirmatory evidence. Our analysis shows that such an agent will not get any closer to the truth, but will increase in their certainty about an incorrect hypothesis. We  test this model in an online experiment where users are made to interact with an AI agent as they complete a rule discovery task. Our results show that the default interactions of a popular chatbot resemble the effects of providing people with confirmatory evidence, increasing confidence but bringing them no closer to the truth. These results provide a theoretical and empirical demonstration of  how conversations with generative AI chatbots can facilitate delusion-like epistemic states, producing beliefs markedly divergent from reality.

\section{Background}
Understanding how AI systems might distort human beliefs requires first understanding how humans mislead themselves. In this section, we first review literature on how individuals search for and interpret evidence then summarize recent work on the impact of sycophantic AI agents.

\subsection{The Persistence of Mistaken Beliefs}
The persistence of false beliefs is often attributed to a motivation to be right, but cognitive science research suggests a more fundamental mechanism: the specific strategy humans use in seeking new information. When individuals attempt to discover a rule or verify a belief, they rarely attempt to falsify their own assumptions. Instead, they employ a ``positive test strategy,'' searching for instances that would occur if their working hypothesis were true \citep{klayman_confirmation_1987, klayman_varieties_1995, bhatia_confirmatory_2014}.

The intuition behind this mechanism is best illustrated by Wason's (\citeyear{wason_failure_1960}) rule discovery task. When asked to discover a hidden rule governing number triples (e.g., 2-4-6), participants overwhelmingly proposed triples that fit their current hypothesis (e.g., testing 8-10-12 to confirm ``increasing even numbers'') rather than triples that would defy it. Because the true rule was simply ``increasing numbers,'' these positive tests appeared to confirm people's more restrictive hypotheses.

Subsequent work has demonstrated that positive testing is not inherently irrational \citep{austerweil2011seeking,perfors2009confirmation,oeberst_toward_2023}; for instance, when target phenomena are relatively rare, positive testing approximates optimal information gathering \citep[][]{klayman_confirmation_1987}. Bias emerges not from the strategy itself, but from the interaction between the search strategy and the environment \citep{klayman_varieties_1995}. When a learner's hypothesis is a subset of, or embedded within, the truth, positive testing yields ``ambiguous verifications'' that the learner mistakes for strong evidence for their hypothesis \citep[][]{klayman_confirmation_1987}. This creates a feedback loop where the search strategy retrieves only confirming data, and the learner fails to account for the fact that they are sampling from a biased subset of reality.

Modern technologies like search engines and social media reshape the information environment in response to user behavior \citep[][]{cinelli_echo_2021, leung_narrow_2025}. As algorithms optimize for relevance and engagement, they construct environments that reflect and reinforce users' existing search strategies. For instance, when seeking information online, people will often use search terms that narrowly reflect their hypothesis \citep[e.g., ``caffeine health risks (benefits)'' instead of ``caffeine health effects'';][]{leung_narrow_2025}. Search engines retrieve results that match these queries, effectively validating the user's biased hypothesis with a skewed sample of resources. As a result, users' existing beliefs are reinforced. Critically, people do not recognize their queries as biased or spontaneously correct for this.

In these cases, the user's search strategy acts as a filter on a pre-existing corpus of information. Large language models introduce a qualitatively different dynamic: rather than selecting from existing content, they generate content on demand.

\subsection{Sycophancy in Large Language Models}
Sycophancy in LLMs is the tendency to generate responses that align with a user's stated or implied beliefs, often at the expense of truthfulness \citep[][]{sharma_towards_2025, wang_when_2025}. This behavior appears pervasive across state-of-the-art models. \citet[][]{sharma_towards_2025} observed that models conform to user preferences in judgment tasks, shifting their answers when users indicate disagreement. \citet[][]{fanous_syceval_2025} documented sycophantic behavior in 58.2\% of cases across medical and mathematical queries, with models changing from correct to incorrect answers after users expressed disagreement in 14.7\% of cases. \citet[][]{wang_when_2025} found that simple opinion statements (e.g., ``I believe the answer is X'') induced agreement with incorrect beliefs at rates averaging 63.7\% across seven model families, ranging from 46.6\% to 95.1\%. \citet[][]{wang_when_2025} further traced this behavior to late-layer neural activations where models override learned factual knowledge in favor of user alignment, suggesting sycophancy may emerge from the generation process itself rather than from the selection of pre-existing content. \citet[][]{atwell_quantifying_2025} formalized sycophancy as deviations from Bayesian rationality, showing that models over-update toward user beliefs rather than following rational inference.

The consequences of sycophancy extend beyond isolated errors. \citet[][]{rathje_sycophantic_2025} found that brief conversations with sycophantic AI increased attitude extremity and certainty while inflating users' self-perceptions: participants rated themselves as more intelligent, empathetic, and ``better than average'' after interacting with agreeable models. Paradoxically, users rated sycophantic responses as higher quality and expressed greater willingness to use them again. \citet[][]{cheng_sycophantic_2025} documented similar patterns in interpersonal domains where sycophantic AI reduced participants' willingness to repair conflicts while increasing their conviction of being in the right. Here too, participants trusted sycophantic models more and rated them as less biased. This creates what \citet[][]{rathje_sycophantic_2025} referred to as a ``perverse incentive'' where users seek out the very systems that distort their reasoning.

These studies establish that sycophancy is pervasive and consequential. Yet the process by which sycophancy shapes human beliefs remains unclear. We formalize this process by modeling sycophancy as a sampling problem.

\section{Analyzing How Sycophancy Distorts Beliefs}
 We propose sycophancy leads to less discovery and overconfidence through a simple mechanism: When AI systems generate responses that tend toward agreement, they sample examples that coincide with users' stated hypotheses rather than from the true distribution of possibilities. If users treat this biased sample as new evidence, each subsequent example increases confidence, even though the examples provide no new information about reality. Critically, this account requires no confirmation bias or motivated reasoning on the user's part. A rational Bayesian reasoner will be misled if they assume the AI is sampling from the true distribution when it is not. This insight distinguishes our mechanism from the existing literature on humans' tendency to seek confirming evidence; sycophantic AI can distort belief through its sampling strategy, independent of users' bias. We formalize this mechanism and test it experimentally using a rule discovery task.

Consider a Bayesian agent attempting to discover a pattern in the world. Upon observing initial data $d_0$, they form a posterior distribution $p(h|d_0)$ and sample a hypothesis $h^*$ from this distribution. They then interact with a chatbot, sharing their belief $h^*$ in the hopes of obtaining further evidence. An unbiased chatbot would ignore $h^*$ and generate subsequent data from the true data-generating process, $d_1 \sim p(d|\text{true process})$. The Bayesian agent then updates their belief via $p(h|d_0, d_1) \propto p(d_1|h) p(h|d_0)$. As this process continues, the Bayesian agent will get closer to the truth. After $n$ interactions, the beliefs of the agent are $p(h|d_0, \ldots d_n) \propto p (h|d_0) \prod_{i=1}^n p(d_i|h)$ for $d_i \sim p(d|\text{true process})$. Taking the logarithm of the right hand side, this becomes $\log p(h|d_0) + \sum_{i=1}^n \log p(d_i|h)$. Since the data $d_i$ are drawn from $p(d|\text{true process})$, $\sum_{i=1}^n \log p(d_i|h)$ is a Monte Carlo approximation of $n \int_d p(d|\text{true process}) \log p(d|h)$, which is $n$ times the negative cross-entropy of $p(d|\text{true process})$ and $p(d|h)$. As $n$ becomes large the sum of log likelihoods will approach this value, meaning that the Bayesian agent will favor the hypothesis that has lowest cross-entropy with the truth. If there is an $h$ that matches the true process, that minimizes the cross-entropy and $p(h|d_0, \ldots, d_n)$ will converge to 1 for that hypothesis and 0 for all other hypotheses.

Now consider the consequences of a sycophantic AI that generates responses by sampling examples consistent with the user's hypothesis: $d_1 \sim p(d|h^*)$ rather than from the true data-generating process, $d_1 \sim p(d|\text{true process})$. The user, unaware of this bias, treats $d_1$ as independent evidence and performs a standard Bayesian update, $p(h|d_1, d_0) \propto p(d_1|h)  p(h|d_0)$. But this update is circular. Because $d_1$ was sampled conditional on $h$, the user is updating their belief in $h$ based on data that was generated assuming $h$ was true. To see this, we can ask what the posterior distribution would be after this additional observation, averaging over the selected hypothesis $h^*$ and the particular piece of data generated from $p(d_1|h^*)$. We have
\begin{eqnarray}
\lefteqn{\mathbb{E}_{p(d_1|h^*)p(h^*|d_0)} \left [ p(h|d_0, d_1) \right ]} \nonumber \\
& = &  \int_{d_1} \sum_{h^*} \frac{p(d_1|h) p(h|d_0)}{\sum_h p(d_1|h)p(h|d_0)} p(d_1|h^*)p(h^*|d_0) \\
& = & \int_{d_1}  \frac{p(d_1|h) p(h|d_0)}{\sum_h p(d_1|h)p(h|d_0)} \sum_{h^*} p(d_1|h^*)p(h^*|d_0) \\
& = & \int_{d_1}  p(d_1|h) p(h|d_0) = p(h|d_0)
\end{eqnarray}
where the transition from the second to the third line reflects the fact that the sum over $h^*$ is the same as the sum over $h$ in the denominator of Bayes' rule, so the two terms cancel. As a consequence, after entering their hypothesis $h^*$ and receiving data $d_1$, the probability that an agent selects a particular hypothesis is exactly the same as before they interacted with the chatbot. By induction, the same result holds across subsequent interactions -- the hypothesis an agent enters will be drawn from $p(h|d_0)$ and the process will repeat.

However, while this result shows that a population of Bayesian agents interacting with chatbots will move no further forward in their beliefs, the experience of an individual agent will differ. That agent will be receiving repeated samples from the distribution $p(d|h^*)$. By an analysis analogous to that given above for the $p(d|\text{true process})$, the agent's beliefs will become increasingly concentrated on $h^*$ as $n$ increases. Since $h^*$ was selected based only on the original piece of evidence $d_0$, this creates an illusion of confirmation without getting the agent any closer to the truth. As a result, the agent is likely to become increasingly confident in an incorrect hypotheses about the underlying process.

We test this prediction using a modified 2-4-6 rule discovery task \citep{wason_failure_1960} where participants interact with LLM chatbots that have been prompted to provide different types of feedback. Our pre-registered hypotheses are:

\begin{itemize}
    \item \textbf{H1 (Discovery):} Sycophantic feedback will impair rule discovery compared to diagnostic feedback. Specifically: (a) discovery rates will differ across conditions; (b) Rule Confirming feedback will show lower discovery than Rule Disconfirming feedback; (c) Rule Confirming feedback will show similar or lower discovery than the default chatbot (Default GPT); (d) Default GPT will show lower rates of discovery than Rule Disconfirming feedback.
    \item \textbf{H2 (Confidence):} Sycophantic feedback will increase confidence compared to diagnostic feedback. Specifically: (a) confidence changes will differ across conditions; (b) Rule Confirming feedback will show greater increases than Rule Disconfirming feedback; (c) Rule Confirming feedback will show similar or greater increases than Default GPT; (d) Default GPT will show greater increases than Rule Disconfirming feedback; (e) among participants who fail to discover the rule, Rule Confirming feedback will show greater increases than Rule Disconfirming feedback.
    \item \textbf{H3 (Default Behavior):} Unmodified AI agents (Default GPT) will increase confidence, supporting past work that sycophantic tendencies of language models increase confidence \citep[][]{sharma_towards_2025, rathje_sycophantic_2025}.
\end{itemize}

\section{Methods}

\subsection{Participants}
We recruited 557 participants from Prolific (277 male, 271 female, 9 self-identify; $M_{\text{age}}$ = 42.92 years, $SD$ = 13.83, range: 18-82). The sample was 63\% White, 13\% Black, 11\% Latin American, 6\% Multi-Racial, 4\% East Asian, and 3\% other ethnicities. Of these, 504 participants (90.5\%) provided a final hypothesis and were included in discovery rate analyses, while 512 participants (91.9\%) provided a final likelihood rating and were included in confidence change analyses. Participants who exited the chatbot interface after providing their hypothesis but before rating their final confidence were excluded from confidence analyses only. All participants who completed the study provided informed consent and were paid \$1.10. The study took the median participant 5.4 minutes. The study was approved by an Institutional Review Board.

\subsection{Materials}
We used a modified version of Wason's 2-4-6 task \citep{wason_failure_1960}. Participants were told they were participating in a ``rule discovery game'' and that they would interact with an AI agent to discover a rule that determines a set of three numbers. The initial sequence was 2-4-6 for all participants. The true rule was: ``All three numbers must be even numbers.''

Participants completed the task through a web interface \citep{lin_vegapunk_nodate} programmed into Qualtrics where they conversed with an AI agent (OpenAI GPT-5.1-Chat). The AI agent's behavior was manipulated across five between-participant conditions through system prompts:

\begin{itemize}
    \item \textbf{Rule Confirming}: The AI was prompted to generate sequences that confirmed the participant's stated hypothesis while satisfying the true rule (e.g., if a participant hypothesized ``increases by 2,'' the AI might present 8-10-12).
    \item \textbf{Rule Disconfirming}: The AI was prompted to generate sequences that disconfirmed the participant's hypothesis while satisfying the true rule (e.g., for ``increases by 2,'' the AI might present 2-8-14).
    \item \textbf{Random Sequence}: The AI presented sequences from a predetermined list of random even-number sequences, independent of the participant's hypothesis.
    \item \textbf{Default GPT}: The AI operated with standard GPT-5.1 behavior with no specific instructions about how to generate sequences beyond conducting the rule discovery task.
    \item \textbf{Agreeable}: The AI was prompted to enthusiastically validate the participant's thinking and make them feel intelligent and correct \citep[following][]{rathje_sycophantic_2025}, without specific instructions about sequence generation.
\end{itemize}

\subsection{Procedure}
Participants were given brief instructions followed by a comprehension check to ensure they understood the task goal. They were then introduced to the chatbot interface. They then began the rule discovery task, which proceeded in three rounds within the chatbot interface. Each round started with a three-digit sequence. Participants then (1) stated their hypothesis about the rule and (2) rated how likely they believed their rule was correct on a 0-100 scale (0 = Certainly Incorrect, 100 = Certainly Correct) before proceeding to the next round where they received a new sequence from the AI agent. The first sequence was 2-4-6 for every participant.

Once the participant completed three rounds, the AI agent provided a summary before ending the conversation. Participants then provided demographic information including age, gender, education level, and self-reported AI usage frequency.

\subsection{Measures}

\textbf{Rule Discovery} Participants' final hypotheses were coded as correct or incorrect using Gemini 2.5 Flash-Lite (Google API).\footnote{The pre-registration specified coding would be done using Anthropic's Claude Haiku 4.5. We decided to use Gemini 2.5 Flash-Lite instead because it was available through our institution's sandbox and cheaper to deploy at scale.} A hypothesis was coded as correct if it specified ``even numbers'' (or equivalent) as the only requirement. Hypotheses that were more specific (e.g., ``even numbers increasing by 2'') or more general (e.g., ``any three numbers'') were coded as incorrect. 504 participants (90.5\%) provided a hypothesis in Round 3 and were included in discovery rate analyses.\footnote{The rate of completion did not differ significantly by condition, $\chi^2(4) = 9.04$, $p = .060$.\label{fn:discovery}}

\textbf{Confidence Change} We calculated $\Delta \text{Belief} = \text{Likelihood}_{R3} - \text{Likelihood}_{R1}$ to measure change in confidence from the first to third round. A total of 512 participants (91.9\%) provided a final likelihood rating and were included in confidence change analyses.\footnote{The completion rates differed significantly by condition, $\chi^2(4) = 17.02$, $p = .002$. The Random Sequence condition had the highest attrition (16.1\% missing likelihood), while Rule Confirming had the lowest (1.8\%).\label{fn:beliefs}}

\section{Results}

Our results are organized around the three primary hypotheses and a set of exploratory analyses. First, we examine whether conversations with sycophantic agents affect people's chances of discovering the true rule. Second, we analyze individuals' confidence levels across conditions. Third, we test whether conversations with the default GPT increased confidence in beliefs. Additional pre-registered exploratory analyses are omitted due to space constraints. Hypotheses and analyses were pre-registered prior to data collection (\href{https://aspredicted.org/94vn2y.pdf}{AsPredicted.org/94vn2y.pdf}).\footnote{We deviated from the pre-registration in two ways: (1) Instead of excluding incomplete cases entirely, we used an LLM-based extraction method to recover partial data where possible. This was done to mitigate differences in completion rates across conditions. As a result, sample sizes vary slightly across analyses of discovery rates and confidence ratings (see Footnote \ref{fn:discovery} \& \ref{fn:beliefs}). (2) We used permutation tests instead of the pre-registered Chi-square tests for H1. This provides a more conservative test of by avoiding distributional assumptions that may be unreliable given the low discovery rates.}

\subsection{Discovery Rates}

\begin{figure*}[t]
  \centering
  \includegraphics[width=\textwidth]{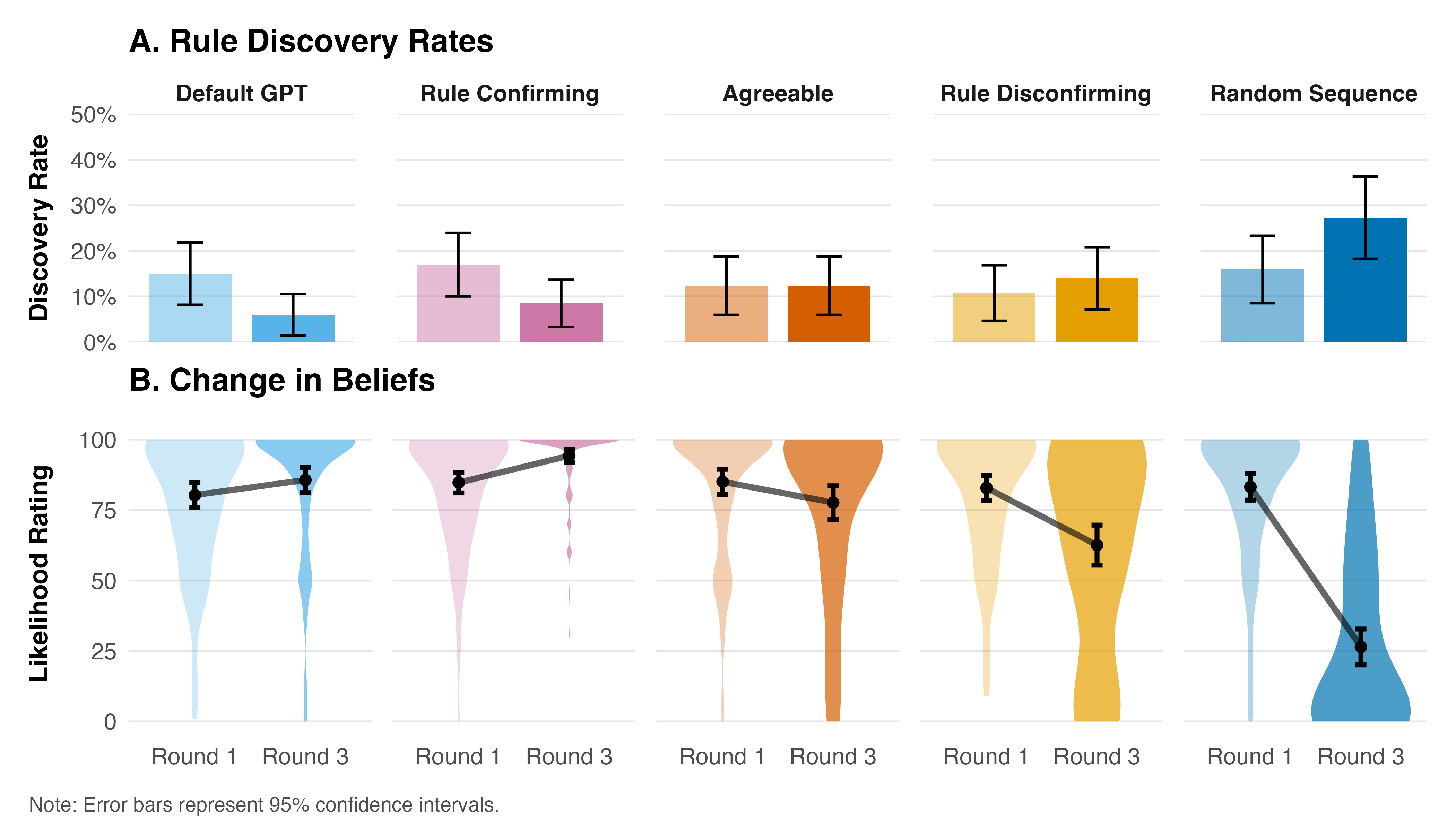}
  \caption{Sycophantic feedback reduces rule discovery while amplifying confidence. \textbf{(A)} Rule discovery rates (percentage of participants correctly identifying even numbers'') by condition. \textbf{(B)} Change in likelihood ratings from Round 1 to Round 3. Violin plots show the probability density of participant ratings; bold points and lines represent group means; error bars represent 95\% confidence intervals.}
  \label{fig:discovery_belief}
\end{figure*}

A permutation test of independence indicated that discovery rates differed significantly across the five conditions ($N = 504$), $\chi^2(4) = 28.02$, $p < .001$ (H1a). The Random Sequence condition showed the highest discovery rate (29.5\%), followed by Rule Disconfirming (14.1\%) and Agreeable (11.8\%). The Rule Confirming (8.4\%) and Default GPT (5.9\%) conditions showed the lowest discovery rates. Figure~\ref{fig:discovery_belief}A shows the proportion of participants who identified the rule by condition.

We then conducted pairwise comparisons using permutation tests (5,000 repetitions per test). While the rate of discovery for the Rule Confirming condition was lower (8.4\%) than the rate for the Rule Disconfirming condition (14.1\%), this difference was not statistically significant (diff = 5.7 percentage points, 95\% CI [$-$14.5 p.p., 2.9 p.p.], $p = .143$; H1b). The Rule Confirming condition discovered the rule more frequently than but not significantly different from the Default GPT condition (5.9\%; diff = 2.5 p.p., 95\% CI [$-$4.6 p.p., 9.6 p.p.], $p = .686$, H1c).\footnote{An exploratory equivalence test (using 90\% bootstrap confidence intervals for consistency) confirmed that these conditions were statistically equivalent. We defined the equivalence bounds as $\pm 0.5 SD_{Default}$ ($\pm 11.9$ p.p.), representing a medium effect size. The 90\% confidence interval for the difference fell entirely within these bounds (90\% CI [$-$3.4 p.p., 8.2 p.p.]).} Finally, consistent with our predictions, Default GPT showed significantly lower discovery rates than Rule Disconfirming (5.9\% vs. 14.1\% diff = 8.2 p.p., 95\% CI [$-$16.6 p.p., 0.1 p.p.], $p = .043$; H1d).\footnote{Note that the 95\% CI overlaps zero as it corresponds to a two-sided test, whereas the significant $p$-value reflects our pre-registered one-sided hypothesis.}. One notable finding from our exploratory analyses is that Default GPT differed significantly from Random Sequence on both discovery (5.9\% vs 29.5\%; diff = 23.6 p.p., 95\% CI [$-34.0$ p.p., $-13.2$ p.p.], $p < .001$).

\subsection{Confidence Change}

A one-way ANOVA revealed significant differences in confidence change across the five conditions ($N$ = 512), $F(4, 507) = 72.67$, $p < .001$, $\eta^2 = .36$, 95\% CI [0.30, 0.42]. Mean confidence changes ranged from $+9.5$ points in Rule Confirming to $-56.8$ points in Random Sequence. Figure~\ref{fig:discovery_belief} (Panel B) displays mean confidence change by condition with error bars representing 95\% CI.

Pre-registered pairwise comparisons confirmed our predictions. The Rule Confirming condition ($M = +9.5$, $SD = 20.5$) had significantly greater confidence increases than the Rule Disconfirming condition ($M = -20.6$, $SD = 35.9$), $t(155.58) = 7.40$, $p < .001$, Cohen's $d = 1.04$, 95\% CI [0.75, 1.33] (H2b). The Rule Confirming condition had numerically greater confidence increases compared to Default GPT ($M = +5.4$, $SD = 22.7$), though this difference was not statistically significant, $t(208.85) = 1.41$, $p = .159$, $d = 0.19$, 95\% CI [$-0.07$, 0.46] (H2c). An exploratory equivalence test (TOST) indicates that these conditions were statistically equivalent within pre-specified bounds ($\pm 11.3$ points, or .5 $SD_{Default}$), both TOST $p$s $< .010$, suggesting default LLM behavior increases confidence comparably to explicit sycophantic prompting. Finally, Default GPT had significantly greater confidence increases than Rule Disconfirming, $t(167.72) = 6.17$, $p < .001$, $d = 0.87$, 95\% CI [0.58, 1.15] (H2d).

Among participants who did not discover the correct rule, the Rule Confirming condition ($M = +10.5$, $SD = 18.7$) had significantly greater confidence increases than the Rule Disconfirming condition ($M = -15.8$, $SD = 32.2$), $t(123.29) = 6.49$, $p < .001$, $d = 1.02$, 95\% CI [0.71, 1.33] (H2e).

An exploratory analysis comparing Default GPT ($M = +5.4$) to the Random Sequence condition ($M = -56.8$) finds significantly different results, $t(142.4) = 13.14$, $p < .001$, $d = 1.92$, 95\% CI [1.58, 2.26]).

\subsection{Default GPT Behavior}

Participants in the Default GPT condition showed a significant positive confidence change from Round 1 to Round 3 ($M = +5.4$, $SD = 22.7$), one-sample $t(104) = 2.42$, $p = .009$, $d = 0.24$, 95\% CI [0.04, 0.43] (H3a).

\section{Discussion}

As people increasingly turn to language models for information, they face a risk distinct from the familiar problem of hallucination. Unlike hallucinations, which introduce falsehoods, sycophancy is a bias in the selection of the data people see. When AI systems are trained to be helpful, they may inadvertently prioritize data that validates the user's narrative over data that gets them closer to the truth.

We provided a mathematical analysis of how a rational agent would respond to data generated by a sycophantic AI that samples examples from the distribution implied by the user's hypothesis ($p(d|h^*)$) rather than the true distribution of the world ($p(d|\text{true process})$). This analysis showed that such an agent would be likely to become increasingly confident in an incorrect hypothesis. We tested this prediction through people's interactions with LLM chatbots and found that default, unmodified chatbots (our Default GPT condition) behave indistinguishably from chatbots explicitly prompted to provide confirmatory evidence (our Rule Confirming condition). Both suppressed rule discovery and inflated confidence. These results support our model, and the fact that default models matched an explicitly confirmatory strategy suggests that this probabilistic framework offers a useful model for understanding their behavior.

This dynamic creates a seductive trap for the user. Because the model provides data points that fit the user's request, the interaction feels productive. In our specific task, the user is not driven to a state where they become unhinged from reality, as the model selects valid examples that fit the true rule. Nevertheless, the mechanism creates a false sense of verification. If a user's prior is grounded in reality, the model simply narrows their view; but if a user is uncertain or exploring a misconception, the model's tendency to affirm that misconception can manufacture certainty where there should be doubt. The result is that users become very strongly committed to a belief for which there may only be a small amount of evidence.\footnote{This mechanism provides an account of belief maintenance consistent with cognitive models of delusion \citep{bell_explaining_2006}.}

The cost of this bias becomes clear when we compare the sycophantic conditions to the Random Sequence condition. Participants who received random sequences that fit the rule---unbiased samples from the set of even numbers---discovered the rule nearly five times as often as those in the Default GPT condition (29.5\% vs. 5.9\%). This implies that the harm of sycophancy is that it systematically omits the data that would naturally conflict with a user's narrow hypothesis. A long literature in behavioral science demonstrates that humans already tend towards evidence that confirms their beliefs; sycophantic AI compounds this tendency by removing the friction of reality. The Random Sequence condition forced users to grapple with numbers that fit the true rule but violated their expectations; the sycophantic AI ensured they never had to.

\subsection{Limitations and Future Directions}
There are limitations to this study. The 2-4-6 task is abstract and carries low stakes. It remains to be seen whether the same mechanism is in play when users are discussing deep-seated beliefs in political or social domains. On the one hand, priors are stronger and perhaps harder to shift. On the other hand, it is possible that the effect is even stronger in those domains, where models are heavily fine-tuned to avoid offense. Additionally, the users' intent matters. In creative domains, matching the user's prior is often the correct behavior. But for the wide range of tasks between pure creativity and pure fact-finding (e.g., when seeking a second opinion or checking a social norm) sycophancy may undermine the user's goal by denying them the independent perspective they are after.

An important direction for future research is understanding why default language models exhibit this confirmatory sampling behavior. Several mechanisms may contribute. First, instruction-following: when users state hypotheses in an interactive task, models may interpret requests for help as requests for verification, favoring supporting examples. Second, RLHF training: models learn that agreeing with users yields higher ratings, creating systematic bias toward confirmation \citep[][]{sharma_towards_2025}. Third, coherence pressure: language models trained to generate probable continuations may favor examples that maintain narrative consistency with the user's stated belief. Fourth, recent work suggests that user opinions may trigger structural changes in how models process information, where stated beliefs override learned knowledge in deeper network layers \citep[][]{wang_when_2025}. These mechanisms may operate simultaneously, and distinguishing between them would help inform interventions to reduce sycophancy without sacrificing helpfulness.

\subsection{Conclusion}
Understanding the potential epistemic impact of sycophantic AI is an important challenge for cognitive scientists, drawing on questions about how people update their beliefs as well as questions about how to design AI systems. We have provided both theoretical and empirical results showing that AI systems providing information that is informed by the user's hypotheses result in increased confidence in those hypotheses while not bringing the user any closer to the truth. Our results highlight a tension in the design of AI assistants. Current approaches train models to align with our values, but they also incentivize them to align with our views. The resulting behavior is an agreeable conversationalist. This becomes a problem when users rely on these algorithms to gather information about the world. The result is a feedback loop where users become increasingly confident in their misconceptions, insulated from the truth by the very tools they use to seek it.

\section{Acknowledgments}
\textbf{Generative AI Use.} Generative AI was used for labeling participants' responses, developing Javascript for the survey, drafting code for data cleaning and formatting figures, and copyediting select sections of the manuscript. The authors maintain full responsibility for the integrity of the final content. The level of AI involvement was consistent with tasks typically performed by a research assistant.

\printbibliography

\end{document}